\documentclass[sigconf,nonacm]{acmart}

\renewcommand\footnotetextcopyrightpermission[1]{}

\usepackage{booktabs}
\usepackage{enumitem}

\title[RTG Control Diagnostics for DT Recommendation]{Auditing Return Conditioning as a Control Knob: An Offline Diagnostic for Decision Transformer Recommendation}

\author{Jingyu Wang}
\affiliation{%
  \institution{Independent Researcher}
  \country{}
}
\email{}

\begin{abstract}
Offline return-to-go (RTG) sweeps can test whether a recommender conditioned on
return is controllable, but the intervention is rarely audited. Rewriting every
historical RTG token creates an increasingly synthetic context, while rewriting
only the current token is more local. We test this distinction in an offline
setting with a fixed window. On MovieLens 25M and MyAnimeList 2020 (MAL), we evaluate a Decision
Transformer using an RTG locality ladder, a control without RTG, a logged
match and score reward check, and a within-trajectory shuffled RTG ablation. On MovieLens, a
\(K=20\) intervention that covers the full context, applied only to real context
positions, shifts the share of Crime predictions by
\(+23.61\pm2.96\) percentage points from the validation 5th to 95th percentile,
whereas changing only the current slot shifts it by
\(+1.77\pm1.17\) points. The shuffled RTG model largely removes this response
(\(+2.08\pm1.20\) points at \(K=20\)). On MAL, the same protocol
does not produce a Drama response: \(K=20\) changes Drama by
\(-0.03\pm0.07\) points, and \(K=1\) by \(-0.01\pm0.01\). Genre prediction
accuracy is numerically close across real RTG, no RTG, and shuffled RTG, and at
\(K=1\) logged match rates and matched ratings change little. Because dataset and
focus-genre selection were exploratory, these magnitudes are descriptive; the
cross-diagnostic pattern across locality, shuffled RTG, and the null result on
MAL does not establish reward control. We propose four checks:
intervention locality, a no-RTG baseline, a reward check, and RTG-content ablation.
\end{abstract}



\begin{document}
\maketitle

\section{Motivation}

Decision Transformers (DTs) act from past interaction tokens and a requested
return-to-go (RTG)~\cite{chen2021decision}. In recommendation, this suggests
that a higher target return may steer recommendations toward longer-term
objectives. Several recommenders use this
idea~\cite{chen2024edt4rec,gao2025tadt,ocejo2025notification}.
MocDT, for example, conditions generated item sequences on specified
multi-objective targets~\cite{gao2025mocdt}. Some studies
provide interactive evidence. EDT4Rec uses an online simulator~\cite{chen2024edt4rec},
while TADT-CSA reports simulation and live A/B test results~\cite{gao2025tadt}.
These studies show policy improvement but do not isolate the
effect of changing the return prompt. We study the static sweep used
when interaction is unavailable. The original DT evaluation, for example,
compares achieved and target returns~\cite{chen2021decision}. What does such an
offline sweep alone establish?

We audit an evaluation where a DT predicts one action from a context of 20
logged interactions. Historical RTGs use future logged behavior. Replacing all of them with
one target constructs a synthetic history. Replacing only the current RTG
changes less of the context, but still mixes a requested target with unavailable
hindsight RTGs. This gap between logged data and intervention has motivated
counterfactual evaluation and policy evaluation from logged data in
recommendation~\cite{schnabel2016recommendations}, and the offline RL
literature has documented closely related failure modes for supervised learning
conditioned on return~\cite{brandfonbrener2022rcsl,paster2022luck,emmons2022rvs}.
We ask whether an offline RTG response reflects local control or sensitivity to a synthetic rewritten context.

Our RTG locality ladder replaces the logged RTG at the trailing \(K\) real
context positions with one target value; states, actions, and padding remain
unchanged.
\(K=1\) is the closest static perturbation to an intervention at the current
decision. \(K=20\) is our endpoint that covers the full context. Varying \(K\)
tests whether the response requires an
increasingly synthetic context. We pair it with three checks. We ask whether RTG
content improves prediction over a baseline without RTG, whether distribution
shifts improve matched rewards, and whether per-position RTG alignment matters during training.
A probe that predicts RTG from state provides a descriptive redundancy check. Concurrent
work addresses related problems: Decoupled DT removes historical RTGs as
unnecessary for action prediction~\cite{wang2026decoupling}, while Q-ALIGN
trains for agreement between requested RTG and policy value on D4RL
benchmarks~\cite{yang2026qalign}. Our ladder instead audits what historical RTG
rewriting establishes in logged recommendation windows without policy rollouts.

\section{Setup}
\label{sec:setup}

Each user trajectory is represented as \((s_t,a_t,r_t,R_t)\). The state
\(s_t\) summarizes interactions strictly before \(t\): a running histogram of
genre preferences plus mean rating and logarithmically normalized count. The
action \(a_t\) is the logged item's primary genre.
The reward \(r_t\) is rating/5 on MovieLens and rating/10 on MAL. With positions indexed
from 0 to \(T-1\), the RTG is a fixed sum over a forward window of length \(W=20\),
\[
R_t=\sum_{i=t}^{t+W-1} r_i,\qquad t+W\le T,
\]
not an episodic return through the end of a trajectory. Positions without a
complete future window of \(W\) interactions are excluded. Our formulation
treats the end of a dataset record as censoring, not as a terminal state with
zero reward. Related recommendation objectives include delayed
conversion~\cite{chapelle2014delayed} and notification objectives with multiple
rewards~\cite{ocejo2025notification}. Here, the RTG is the sum of normalized
ratings over 20 interactions starting at \(t\). The model and probe receive
\(R_t/20\). Sweep targets are reported on the unscaled summed-return axis. We use interactions
rather than calendar time because MAL has no timestamps and this choice aligns
the reward window with the model context. The context length is also \(20\).
We keep these terms separate in the text, using \(W\) for the reward window and
``\(K\)'' for the number of overwritten context positions. Table~\ref{tab:scope}
summarizes the evaluation scope: MovieLens 25M~\cite{grouplens2019movielens25m} with 18 genre actions and
MyAnimeList Database 2020 (MAL)~\cite{hernan2020myanimelist} with 15 genre actions. Genre
actions make distribution shifts measurable, but limit the conclusion to genre
prediction. To form one action from multi-genre items, MovieLens chooses the
least frequent genre to limit dominance by common tags; MAL uses a fixed
niche-to-broad priority. These mappings can shape which genre responds;
alternative mappings remain future work. MAL has no timestamps, so each user's order
follows dataset rows after filtering. It provides a replication that is
sensitive to ordering and cannot support claims about temporal behavior over a long horizon.

Dataset selection was exploratory, not preregistered. We screened seven alternatives
for sufficient trajectories, a usable action space and ordering, and a catalog
distinct from MovieLens. Five failed these feasibility checks. Amazon CDs reached
exploratory modeling but was not retained as a replication dataset: its action
distribution was concentrated (Pop comprised 30.9\% of actions), and its
preliminary RTG sweep was null. Because this outcome-aware selection can favor
datasets with detectable effects, we treat MovieLens as exploratory rather than
confirmatory. MAL was retained as the cross-domain replication. Its final
result is also null.

The observed genre is a proxy action. These datasets record consumed and rated
items, not recommendation exposures from a known policy. Our experiments
diagnose the next genre in a user's history, not the value of a recommender policy.

The primary protocol predicts the next action from the current state. The model
sees \(s_t\) and \(R_t\), but the current action is replaced with a PAD token:
\[
\{s_i,a_i,R_i\}_{i<t}\cup(s_t,\mathrm{PAD},R_t)\rightarrow a_t.
\]
The DT has 3 transformer layers, hidden size 128, 4 heads, and context length
20. DT-no-RTG zeros the RTG embedding while keeping the architecture. The
shuffled RTG control keeps the RTG channel active but permutes RTG values within
each user's training trajectory. This breaks alignment between a position and
its own future return while preserving that user's RTG multiset and user-level
correlations. SASRec is included as a sequence model
that uses attention and has no RTG input~\cite{kang2018sasrec}.

Logged \(R_t\) includes future outcomes and the current rating \(r_t\) for the
target item. Thus, accuracy with real RTG is diagnostic rather than deployable.
We test control by overwriting RTG tokens at evaluation time.

\begin{table}[t]
  \centering
  \scriptsize
  \setlength{\tabcolsep}{1.7pt}
  \caption{Experimental scope. Filtered counts precede the minimum-history
  requirement (MovieLens: 10 and MAL: 15). Effective test counts additionally
  require complete \(W=20\) RTGs.}
  \label{tab:scope}
  \begin{tabular}{lrrrrr}
    \toprule
    Dataset & Filtered users & Filtered events & Test users & Test contexts & Actions \\
    \midrule
    MovieLens & 162K & 25.0M & 15,805 & 2.18M & 18 \\
    MAL & 309K & 49.3M & 25,819 & 4.36M & 15 \\
    \bottomrule
  \end{tabular}
\end{table}

Two details make the setup conservative. First, the model receives the current
state and RTG while only the unknown action is masked. Second, shuffled RTG keeps
the input channel and each user's value distribution, but removes the alignment
between a position and its future return. A gain that depends on RTG should decrease.

\section{Diagnostics and Results}

Unless noted, \(x\pm y\) is the mean and sample standard deviation over three
training seeds, not a confidence interval. Within each seed, 95\% bootstrap
percentile intervals use 1,000 resamples at the user level. Metrics pool
eligible contexts, so users contribute in proportion to eligible history
length. Resampling handles dependence within each user but still weights each
context equally.

\subsection{Intervention Locality}

\begin{figure}[b]
  \centering
  \includegraphics[width=\linewidth]{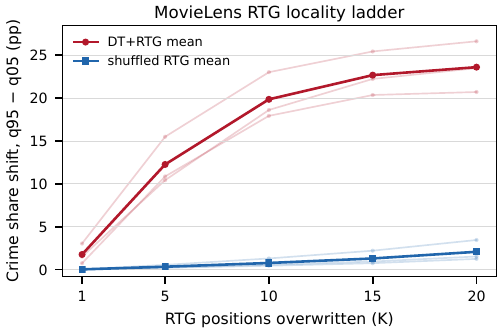}
  \Description{Line plot showing MovieLens Crime share shift as the number of overwritten RTG positions increases from 1 to 20. The mean shift with real RTG rises from about 2 percentage points at K=1 to about 24 points at K=20, while shuffled RTG remains near zero to small.}
  \caption{MovieLens RTG locality ladder. Sweeping RTG across the full context
  produces a large Crime shift, but changing one slot produces a small shift
  that varies across seeds. Shuffled RTG largely removes the response. Faint
  lines are seeds. Bold lines are means.}
  \label{fig:ladder}
\end{figure}

For each sampled test context, we sweep target RTG from the validation
5th to 95th percentile and overwrite only the trailing \(K\) RTG positions.
The final test ranges are from 9.70 to 17.70 on MovieLens and from 11.80 to
18.20 on MAL.
For every reported shift, we evaluate the same test contexts at both endpoints
and compute \(\Delta=\mathrm{metric}(R_{q95})-\mathrm{metric}(R_{q05})\), where
the targets are fixed from the validation RTG distribution.
Figure~\ref{fig:ladder} and Table~\ref{tab:ladder} show that the MovieLens
response depends sharply on \(K\). Changing the full context shifts Crime by
\(+23.61\pm2.96\) percentage points. Changing one slot shifts it by only
\(+1.77\pm1.17\) points and varies across seeds. The shuffled model keeps the
RTG channel but breaks positional content, yielding \(+2.08\pm1.20\) points at
\(K=20\) and \(+0.03\pm0.00\) at \(K=1\). Validation yields
\(+22.26\pm2.90\) and \(+1.59\pm1.00\), respectively, confirming the pattern
across splits. For seeds 42, 0, and 1, the user-level cluster-bootstrap 95\%
intervals span 2.63--3.46, 0.66--0.84, and 1.36--1.68pp at \(K=1\), whereas at
\(K=20\), they span 24.31--29.09, 18.95--22.56, and 21.43--25.66pp.
Crime was chosen in exploratory validation, not preregistered. Across all test
genres, it has the largest absolute response at both endpoints. Sci-Fi is next,
shifting by \(-1.01\pm0.55\)pp at \(K=1\) and
\(-15.65\pm0.87\)pp at \(K=20\).

\begin{table}[t]
  \centering
  \small
  \setlength{\tabcolsep}{4pt}
  \caption{MovieLens Crime share shift, validation q95 target minus q05 target,
  across seeds on the test split.}
  \label{tab:ladder}
  \begin{tabular}{lrrrr}
    \toprule
    \(K\) & s42 & s0 & s1 & mean \(\pm\) std \\
    \midrule
    1  & 3.05  & 0.75  & 1.52  & \(1.77 \pm 1.17\) \\
    5  & 15.49 & 10.86 & 10.43 & \(12.26 \pm 2.81\) \\
    10 & 23.02 & 17.93 & 18.61 & \(19.85 \pm 2.76\) \\
    15 & 25.44 & 20.36 & 22.23 & \(22.68 \pm 2.57\) \\
    20 & 26.62 & 20.71 & 23.49 & \(23.61 \pm 2.96\) \\
    \bottomrule
  \end{tabular}
\end{table}

MAL gives a different caution. Table~\ref{tab:mal} shows no Drama lever under
the current state protocol with PAD masking. Across three seeds, \(K=1\)
changes Drama by \(-0.01\pm0.01\)pp and \(K=20\) by
\(-0.03\pm0.07\)pp. Validation is likewise null, making this a null steering
result rather than local control with a failed reward check. A post hoc audit
over all 15 MAL genres finds a one-slot maximum of only \(0.024\pm0.017\)pp for
Slice of Life and a full-context maximum of only \(0.259\pm0.093\)pp for
Supernatural. The largest difference between real and
shuffled RTG over the full context and all MAL genres is \(0.147\)pp. These
numbers are about two orders of magnitude smaller than the MovieLens saturated
effect and leave no plausible alternative MAL genre for a positive steering
claim.

\begin{table}[t]
  \centering
  \small
  \setlength{\tabcolsep}{4pt}
  \caption{MAL Drama share shift, validation q95 target minus q05 target,
  across seeds on the test split.}
  \label{tab:mal}
  \begin{tabular}{lrrrr}
    \toprule
    \(K\) & s42 & s0 & s1 & mean \(\pm\) std \\
    \midrule
    1  & -0.01 & -0.01 & 0.00 & \(-0.01 \pm 0.01\) \\
    5  & -0.04 & -0.09 & 0.00 & \(-0.04 \pm 0.04\) \\
    10 & -0.08 & -0.09 & 0.00 & \(-0.06 \pm 0.05\) \\
    15 & -0.10 & -0.07 & 0.00 & \(-0.06 \pm 0.05\) \\
    20 & -0.10 & -0.01 & 0.03 & \(-0.03 \pm 0.07\) \\
    \bottomrule
  \end{tabular}
\end{table}

Adjacent RTGs share most future rewards over \(W\) steps. One changed RTG is
therefore surrounded by hindsight RTGs from the same trajectory. Changing all
RTGs removes this conflict, but creates a synthetic context where every position
claims the same future return. The ladder changes both signal consistency and
distance from logged data. Thus, the gap between \(K=1\) and \(K=20\) shows that
the large response depends on rewriting the full context and cannot estimate
a local decision effect.

\subsection{Reward Check Beyond Distribution Shift}

To separate action changes from reward changes, we use logged test data. When
the model's top prediction matches the logged genre, we record its rating. This
metric is not a counterfactual value estimate. It only tests whether matched
rewards rise with the target. This \(K=1\) check is the closest static approximation
to a current-decision intervention and does not assess synthetic \(K=20\).

\begin{figure}[t]
  \centering
  \includegraphics[width=\linewidth]{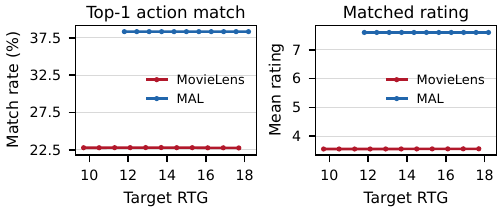}
  \Description{Line chart with two panels for MovieLens and MAL. The match rate of the top prediction and the average matched rating show only small numerical changes as target RTG increases within the observed target range.}
  \caption{Match and score at \(K=1\) for seed 42. Curves change little.
  Table~\ref{tab:match} reports endpoint means across three seeds.}
  \label{fig:match}
\end{figure}

Figure~\ref{fig:match} shows little movement at \(K=1\). Across three seeds, the match rate for the top prediction
changes by \(-0.01\pm0.02\)pp and the matched rating changes by only
\(+0.0041\pm0.0028\) stars (Table~\ref{tab:match}), consistent with the weak
\(K=1\) genre shift in Table~\ref{tab:ladder}. MAL likewise changes little. The
match rate for the top prediction changes by \(0.00\pm0.00\)pp and the
matched rating by \(+0.0001\pm0.0001\) on a 10-point scale. The reward check gives
no indication that the target improves logged ratings. These changes are descriptive:
no paired difference interval or equivalence margin is reported.

\begin{table}[t]
  \centering
  \scriptsize
  \setlength{\tabcolsep}{3pt}
  \caption{Logged match and score at \(K=1\), mean \(\pm\) std over seeds 42,
  0, and 1. Endpoints are validation q05 and q95 targets. Ratings are on the
  native dataset scale.}
  \label{tab:match}
  \begin{tabular}{lrrrr}
    \toprule
    Dataset & Match@q05 & Match@q95 & Rating@q05 & Rating@q95 \\
    \midrule
    MovieLens & \(0.2276\pm0.0002\) & \(0.2275\pm0.0000\) & \(3.5579\pm0.0024\) & \(3.5620\pm0.0033\) \\
    MAL & \(0.3839\pm0.0006\) & \(0.3839\pm0.0006\) & \(7.5984\pm0.0012\) & \(7.5985\pm0.0012\) \\
    \bottomrule
  \end{tabular}
\end{table}

This is weaker than sequential policy evaluation from logged
data~\cite{jiang2016doubly} and remains subject to selection bias in logged
data~\cite{schnabel2016recommendations}. It only scores logged actions, and the
matched set can change with the target. Small changes therefore give no evidence
that the distribution shift improves reward at the user level.

\subsection{Channel Content}

The shuffled RTG control tests whether RTG values carry the effect, motivated by
work on the capabilities and limits of return-conditioned supervised learning
in offline RL~\cite{brandfonbrener2022rcsl,emmons2022rvs}. It preserves the
channel and each user's RTG multiset, but breaks per-position alignment between
the state--action context and its own future return. User-level correlations can remain.
If correct RTG content matters, shuffled RTG should approach no RTG.

\begin{table}[t]
  \centering
  \small
  \setlength{\tabcolsep}{4pt}
  \caption{Test genre-prediction hit rates at 1 (HR@1) and 5 (HR@5), mean
  \(\pm\) sample std over three seeds. Real RTG's advantage over no RTG is small.}
  \label{tab:accuracy}
  \begin{tabular}{llcc}
    \toprule
    Dataset & Model & HR@1 & HR@5 \\
    \midrule
    MovieLens & DT+RTG & \(0.2214\pm0.0002\) & \(0.6343\pm0.0004\) \\
    MovieLens & DT-no-RTG & \(0.2200\pm0.0004\) & \(0.6329\pm0.0002\) \\
    MovieLens & DT shuffled & \(0.2201\pm0.0006\) & \(0.6331\pm0.0001\) \\
    MovieLens & SASRec & \(0.2139\pm0.0003\) & \(0.6241\pm0.0002\) \\
    \midrule
    MAL & DT+RTG & \(0.3822\pm0.0009\) & \(0.7709\pm0.0016\) \\
    MAL & DT-no-RTG & \(0.3824\pm0.0010\) & \(0.7706\pm0.0017\) \\
    MAL & DT shuffled & \(0.3822\pm0.0009\) & \(0.7709\pm0.0017\) \\
    MAL & SASRec & \(0.3730\pm0.0002\) & \(0.7526\pm0.0003\) \\
    \bottomrule
  \end{tabular}
\end{table}

Table~\ref{tab:accuracy} shows that aligned RTG content gives only a small
predictive lift. On MovieLens, real RTG is \(0.14\)pp above the model without
RTG in both HR@1 and HR@5, while shuffled RTG is only \(0.01\)pp above that model
in HR@1. On MAL,
real RTG, no RTG, and shuffled RTG are numerically close. These results are compatible
with weak use of the RTG channel, but they cannot demonstrate reward control.
SASRec confirms that the task is predictable without an RTG channel.

The probe that predicts RTG from state shows that RTG is partly predictable
from state, and on MAL largely so. Ridge regression explains 34.1\% of MovieLens
RTG variance and 73.6\% of MAL variance, while an MLP with one hidden layer explains 50.5\% and
75.5\% (Table~\ref{tab:probe}).

\begin{table}[t]
  \centering
  \small
  \setlength{\tabcolsep}{4pt}
  \caption{Probe that predicts scaled RTG from state on test examples. The
  target is \(R_t/20\), matching the model input, and \(R^2\) is invariant to this
  scaling.}
  \label{tab:probe}
  \begin{tabular}{lrrr}
    \toprule
    Dataset & Ridge \(R^2\) & MLP \(R^2\) & std(\(R/20\)) \\
    \midrule
    MovieLens & 0.341 & 0.505 & 0.124 \\
    MAL & 0.736 & 0.755 & 0.094 \\
    \bottomrule
  \end{tabular}
\end{table}

\begin{table}[t]
  \centering
  \small
  \setlength{\tabcolsep}{3pt}
  \renewcommand{\arraystretch}{0.96}
  \caption{Failure mode each diagnostic probes.}
  \label{tab:diagnostics}
  \begin{tabular}{p{0.25\linewidth}p{0.36\linewidth}p{0.27\linewidth}}
    \toprule
    Diagnostic & Failure mode & Observed here \\
    \midrule
    RTG locality & saturated response read as local control & MovieLens: full \(\gg\) local. MAL max \(<0.3\)pp \\
    \specialrule{0.25pt}{1pt}{1pt}
    No RTG control & predictive lift attributed to RTG without a matched baseline & lift \(\le0.14\)pp at HR@1/5 \\
    \specialrule{0.25pt}{1pt}{1pt}
    Match and score & action shift read as reward gain & small \(K=1\) changes \\
    \specialrule{0.25pt}{1pt}{1pt}
    Shuffled RTG & channel presence conflated with aligned content & MovieLens response collapses. Accuracy \(\approx\) no RTG \\
    \bottomrule
  \end{tabular}
\end{table}

\section{Implications}

The four diagnostics in Table~\ref{tab:diagnostics} support a limited interpretation. In this setting with a
fixed window and genre actions, MovieLens RTG conditioning changes the aggregate
genre mix, but the large effect requires rewriting the full context and is largely removed by
shuffled RTG. On MAL, neither the genre share ladder nor match and score shows a
target response. Correctly aligned RTGs yield at most a small advantage in
prediction accuracy, which is distinct from reward steering. We therefore read the
MovieLens sweep as tracking the aggregate genre distribution rather than demonstrating
rating-based return control for individual users. We read the MAL sweep as a null steering
result. This interpretation draws on aggregate action shares and logged reward
checks without directly estimating policy value.

These findings leave open the effectiveness of recommendation with DTs, control at the
item level, learned control tokens, and canonical autoregressive DT rollouts. Offline RTG
sweeps should be reported as sensitivity analyses unless a local intervention links the
response to reward improvement. The relevant question is whether movement caused by the target is
local, conditioned on the user, and able to improve reward under an intervention the system
could actually use.

\subsection{Relation to Canonical DT Control}

Our analysis covers only a subset of Decision Transformer protocols. In canonical
offline RL, a target return is set at the beginning of an autoregressive rollout
and decremented by realized reward after each action~\cite{chen2021decision}.
Here, prediction is a single forward pass on a logged window of fixed length,
and each historical RTG is a statistic of a future window computed from the log.
The RTG locality ladder is therefore an input sensitivity diagnostic for
studies where interaction is unavailable. A response only under \(K=20\) may
describe a learned correlation rather than a deployable knob for target return. A
response under \(K=1\) still requires a reward check.

\subsection{Scope and Failure Modes}

The genre action space is diagnostic rather than realistic for production. It
makes distribution shifts caused by the target measurable, but deployment at the item level
would add exposure bias, popularity effects, sparse items, and confounds that are
specific to the catalog. Match and score is also limited. It evaluates only
actions present in the test log and cannot recover ratings for unobserved recommendations.
These small changes cannot determine whether a deployed policy would fail.
They show only that this logged diagnostic supplies no positive evidence of reward
improvement.

\subsection{Reproducibility}

One trajectory builder computes states, genre actions, normalized rewards, and
RTGs over \(W\) steps. The same trajectories feed DT, DT-no-RTG, shuffled RTG,
and SASRec. Experiments use the current state protocol with PAD masking in
Section~\ref{sec:setup}, a seeded randomized 80/10/10 split by users for
training, validation, and testing, complete \(W=20\) reward windows, and seeds 42, 0,
and 1. Diagnostic sweeps use seeded
samples for tractability. We use 1,000 test trajectories for the RTG locality
sweep and 2,000 for match and score. Final tables use RTG targets within the
observed support, fixed from the validation 5th to 95th percentiles and evaluated on the
test split. The test users are disjoint from training and validation users, but
the test cohort is not a pristine prospective holdout:
earlier exploratory models had been trained on most users in these public datasets.
Training budgets are 50K steps for MovieLens and 20K for MAL, with checkpoints
selected by validation loss every 5K steps. In
the shuffled RTG control, only the assignment of RTG values within each user's
training trajectory changes. Validation and test RTGs remain real.

\section{Conclusion}

Four checks audit offline return conditioning: current-slot versus full-context
RTG, no-RTG and shuffled-RTG controls, and a logged reward check. MovieLens's
large response requires a synthetic full-context rewrite; its \(K=1\) check
shows no local reward improvement. MAL yields a null steering result. The
contribution is the audit protocol and cross-diagnostic pattern, not any genre
response.

\clearpage
\bibliographystyle{ACM-Reference-Format}
\bibliography{references}

\end{document}